\documentstyle[epsfig]{aipproc}

\begin{document}
\title{LAPW vs. LMTO full-potential simulations\\
and anharmonic dynamics of KNbO$_3$}

\author{A.~V.~Postnikov and G.~Borstel}
\address{University of Osnabr\"uck -- Department of Physics,
D-49069 Osnabr\"uck, Germany}
\date{\centerline{(Received February 4, 1998)}}
\maketitle
\begin{abstract}
With the aim to get an insight in the origin of differences
in the earlier reported calculation results for KNbO$_3$
and to test the recently proposed ``NFP'' implementation of the
full-potential linear muffin-tin orbital (FP-LMTO) method by
M.~Methfessel and M.~van Schilfgaarde,
we perform a comparative study of the
ferroelectric instability in KNbO$_3$ by FP-LMTO and full-potential
linear augmented plane-wave (LAPW) method. It is shown that
a high precision in the description of the charge density
variations over the interstitial region in perovskite
materials is essential; the technical limitations of
the accuracy of charge-density description apparently accounted
for previously reported slight disagreement with the LAPW
results. With more accurate description of the
charge density by sufficiently fine real-space grid,
the results  obtained by both methods became almost identical.

In order to extract additional information (beyond the harmonic
approximation) from the total energy
fit obtainable in total-energy calculations, a scheme is proposed
to solve the multidimensional vibrational Schr\"odinger equation
in the model of non-interacting anharmonic oscillators via
the expansion in hyperspherical harmonics. Preliminary results
are given for the $t_{1u}$ vibrational modes in cubic KNbO$_3$.
\end{abstract}

\section*{Introduction}

KNbO$_3$ is one of benchmark systems for {\it ab initio} analysis
of ferroelectric perovskites. It has been extensively studied
by the whole spectrum of numerical methods from an apparently
ultimately accurate full-potential linear augmented plane waves (LAPW)
\cite{sb92,sin95,sin97,krak95,krak96}
through pseudopotentials \cite{pseudo} and tight-binding {\it ab initio}
schemes \cite{HF} to semiempirical and model-based techniques
\cite{model}.

Modern state-of-art simulations of ferroelectric correlations,
lattice dynamics and phase transitions are dependent on reliable
and accurate description of the total energy as function of
displacements and strain variables. The full-potential linear
muffin-tin-orbitals (FP-LMTO) method \cite{msm} was proven able to provide
a reasonable balance of accuracy and low computational effort
even when applied to supercells of up to 40 atoms
\cite{knbo3,phonon,ortho,willi96}.
However, in sensitive benchmark calculations of $\Gamma$ phonons,
due to certain technical limitations, FP-LMTO provided apparently
lower accuracy for phonon eigenvectors, as compared to
FP-LAPW \cite{phonon,sin97,krak96}.
A new version of the FP-LMTO code by Methfessel
and van Schilfgaarde \cite{nfp100}, in contrast
to the previous one used in our earlier calculations \cite{msm},
is unsensitive to sphere packing and uses much more efficient, albeit more
mathematically involved, basis of augmented ``smooth Hankel functions''
that enables one to drastically reduce the size of diagonalization problem
without loss of accuracy. We compare the results obtained
with the new FP-LMTO and with the WIEN97
implementation of the FP-LAPW method \cite{wien97}, concentrating on the
accuracy and performance.

In order to get an additional insight into lattice dynamics of KNbO$_3$,
we present vibrational frequencies
as calculated quantum-mechanically in the assumption of uncoupled
multidimensional anharmonic oscillators,
based on the total energy data obtained from
first-principles calculations, and give a preliminary estimate of the
lowest vibration frequency within this approach.

\section*{Comparison of calculation methods}

The linear augmented plane-wave (LAPW) and the linear
muffin-tin orbital (LMTO) methods are closely related and originate
in the same work by Andersen \cite{oka75}.
In the modern stage at their development, they share the advantage
of being all-electron methods (in contrast to pseudopotential
ones and to those depending on the frozen-core approximation).
As compared to tight-binding schemes with fixed (e.g. Gaussian-type)
bases, the basis in LAPW and LMTO is optimized in the course
of iterations, as the heads of wavefunctions are recalculated
inside the (arbitrarily) chosen muffin-tin spheres from one
iteration to another.
The tails of basis functions that span the interstitial region
between the spheres are constructed by an augmentation procedure
which matches the numerical solutions inside the MT-spheres
either to the plane waves of different {\bf k} (in LAPW)
or to the spherical Hankel functions of different energy (in LMTO).
This difference accounts for relative advantages and disadvantages
of these two calculation schemes. In LAPW, the number of augmented
plane waves with different {\bf k} can be safely saturated, until
the desirable convergence of results with respect to the completeness
of basis is achieved. This however has its price in terms of computational
effort.

The advantage of LMTO lies in the fact that its basis functions
may be tuned to resemble true atom-centered wavefunctions in a crystal.
This means, in the ideal case, quite efficient and compact basis set
and hence computational speed and the ability to treat larger systems
than is possible with LAPW, given certain amount of computer resources.
The weak point of LMTO is that the saturation of the basis set
is not as straightforward as in the LAPW, because adding more
atom-centered tail functions of certain angular symmetry needs care
to prevent linear dependences within the basis set.
In order to overcome this problem and at the same
time to maintain the LMTO as, ideally, the `minimal basis' method
of competitive accuracy, one can try to experiment with more
sophisticated but hopefully more efficient envelope functions.
As some examples of proposed alternatives to conventional
spherical Hankel functions one can single out the tight-binding LMTO
method \cite{tblmto} or the ``exact muffin-tin orbital theory''
development \cite{exact}, both of them being not yet implemented,
to our knowledge, in workable full-potential total-energy codes.

Recently, yet another extension within the general LMTO formalism
has been proposed and implemented by M.~Methfessel and
M.~van Schilfgaarde and referred to
as the ``NFP'' LMTO code \cite{nfp100}. While in many aspects a development
of the earlier described \cite{msm} and widely used FP-LMTO formalism,
the NFP algorithm incorporates an essential new element,
that is, using the ``smooth'' Hankel functions rather than the ``standard''
ones for the augmentation of numerical radial solutions in the
interstitial region. Whereas the standard spherical Hankel function
satisfies the differential equation
$$
(\Delta + \epsilon)h_0(r) = -4\pi\delta({\bf r})
$$
for $l=0$, the equation for the smooth function contains
the smeared $\delta$ function $g_0(r) = C\exp(-r^2/R_{\mbox{\small sm}}^2)$
as a source term
$$
(\Delta + \epsilon){\tilde h}_0(r) = -4\pi g_0(r)\;,
$$
and the smooth functions of higher orders are generated by applying
a differential operator ${\cal Y}(-\nabla)$, defined by
${\cal Y}({\bf r}) =r^l Y_L$, to the function of the 0th order.
These new envelope functions can be tuned by a proper choice
of the smoothing parameter $R_{\mbox{\small sm}}$ to imitate
the actual shape of the wavefunction in crystal.

Cubic perovskite ferroelectrics provide an excellent benchmark
system for the high-precision total-energy calculation scheme.
Whereas one typically cannot pinpoint any essential disagreement in the
band dispersions calculated by different full-potential schemes,
the energy differences on the $\sim$1 mRy scale related to
ferroelectric instability, soft mode phonon frequencies and
eigenvectors in these materials may be quite differently estimated
by different computation schemes, result in qualitatively different
predictions and thus dramatize the competition between different
numerical approaches. For KNbO$_3$, a number of calculations has
been already done using different methods. Full-potential LAPW
calculations were performed by Singh \cite{sb92,sin95,sin97}
and Krakauer {\it et al} \cite{krak95,krak96}, LMTO calculations by us
\cite{knbo3,phonon,ortho,willi96}. In spite of the overall agreement
(the ability of both methods
to account for the ferroelectric instability, consistent
results for the $\Gamma$-frozen phonon frequencies), some
disagreements prevailed in the description of
ferroelectric instability, as well as in the estimation
of the soft mode phonon eigenvector \cite{phonon,sin97,krak96}.

What makes perovskites generally (and KNbO$_3$ as an example)
a hard test for any computational scheme
relying on a site-centered basis, like LMTO, is their relatively
loose structural packing (if one thinks in terms of nominal
ionic radii). The electron density is unevenly distributed
between compact NbO$_6$ octahedra and intermediate large cavities
which host small K ions. In our earlier LMTO calculations
using the ``old'' LMTO code by Methfessel\cite{msm}, good sphere
packing was essential for accurate integration over the
interstitial region, but could not be guaranteed in a completely
satisfactory way (see, e.g. the discussion in Ref.~\cite{willi96}).
In the present implementation of LMTO, the aspect of good sphere
packing is no more sensitive, therefore we tried to figure out
finally whether the mentioned disagreement with the LAPW results
was due to problems of technical inefficiency (unadequate integration
scheme etc.) or has to do with some basic limitations of the LMTO
formalism.

In our present benchmark calculations, we used the implementation
of the full-potential LAPW method knows as the WIEN97 code
\cite{wien97}. Sphere sizes were chosen as shown in the
Table \ref{tab:setup} (the same for LAPW and LMTO).
The {\bf k}-space integration was performed in an identical way
in both schemes, using the sampling on a mesh of 18
inequivalent points, corresponding to $6\!\times\!6\!\times\!6$
divisions of the full Brillouin zone. This mesh was found to
be sufficiently dense for the estimations of ferroelectric
instability, based on previous experience \cite{sin95}.
The exchange-correlation was treated either in the local density
approximation (LDA) or in the generalized gradient approximation (GGA).

\begin{table}
\caption{NFP-LMTO calculation setup ($\kappa^2$ in Ry,
$R_{\mbox{\tiny MT}, \mbox{\tiny SM}}$ in a.u.)}
\label{tab:setup}
\begin{tabular}{cdd@{\hspace*{1.5cm}}cdd@{\hspace*{1.5cm}}cdd}
\multicolumn{3}{c}{\rule[-1mm]{0mm}{0mm}
K ($R_{\mbox{\tiny MT}}$=1.95)} &
\multicolumn{3}{c}{Nb ($R_{\mbox{\tiny MT}}$=1.85)} &
\multicolumn{3}{c}{ O ($R_{\mbox{\tiny MT}}$=1.55)} \\
\tableline
\rule[-2mm]{0mm}{6mm}
 $n,l$ & $\kappa^2$ & $R_{\mbox{\tiny SM}}$ &
 $n,l$ & $\kappa^2$ & $R_{\mbox{\tiny SM}}$ &
 $n,l$ & $\kappa^2$ & $R_{\mbox{\tiny SM}}$ \\
\tableline
\rule[0mm]{0mm}{4mm}
 $4s$  & $-$0.5 & 3.0 & $5s$ & $-$0.1 & 2.0 & $2s$    & $-$0.5  & 0.79 \\
 $3p$  & $-$0.5 & 1.3 & $4p$ & $-$1.9 & 0.9 & $2p$    & $-$0.15 & 0.71 \\
 $3d$  & $-$0.2 & 3.4 & $4d$ & $-$0.5 & 1.2 & $s,p,d$ & $-$0.2  & 2.0  \\
\rule[-1mm]{0mm}{0mm}
 $s,p$ & $-$0.2 & 2.0 & $s,p,d$ & $-$1.0 & 1.5 \\
\end{tabular}
\end{table}

In the construction of the LMTO setup, the usual procedure is
to optimize the possibly minimal basis, using the freedom in the
choice of basis parameters, and then to expand the basis
in order to ensure its sufficient completeness.
In the NFT formalism, the quality of the basis depends
on both energies of the smooth Hankel functions and their
smoothing radii, the latter apparently having more pronounced
effect. In contrast to earlier FP-LMTO scheme \cite{msm} which
favored the basis functions with at least three tail energies
per orbital for sufficient accuracy, the NFP provides a reasonable
description of the valence band states with augmenting just one
smooth Hankel function to each; the setup is then refined
by adding some other tail functions.
The calculation setup we used in the present LMTO calculation is shown
in Table \ref{tab:setup}. It includes 70 basis functions, that
can be either somehow more extended, or reduced, to get the desirable
compromise between the accuracy and performance. For comparison,
the basis size for a LAPW calculation of comparable accuracy
should include at least $\sim$800 augmented plane waves.

The present version of NFP incorporates
only the LDA treatment of the exchange/correlation.
Another technical drawback is the impossibility to treat
the states with different principal quantum number and the same
orbital quantum number within the valence band. For KNbO$_3$,
we had to include Nb$4p$ states and neglect
Nb$5p$ in the valence panel. This seems to be an acceptable
compromise, however. More discussion related to the
LMTO setup may be found in Ref. \cite{knbo3}.

\section*{Calculation results and discussion}

\begin{figure}[b!] 
\centerline{\epsfig{file=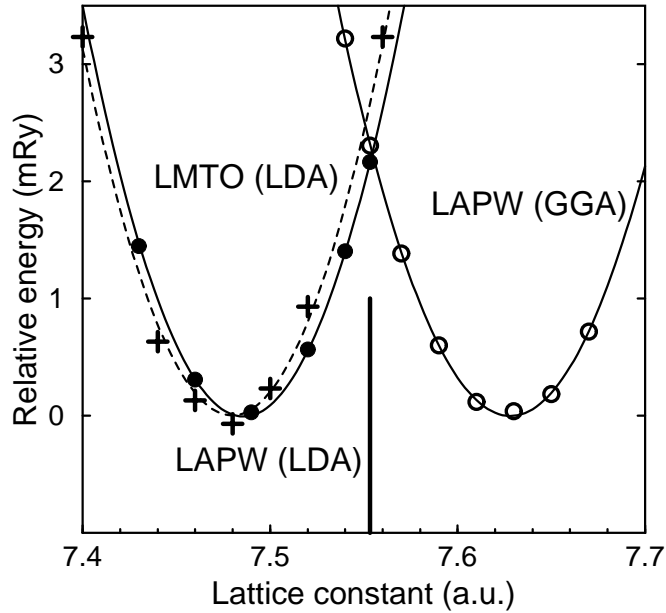,height=8.0cm}}
\vspace{10pt}
\caption{Total energy difference (with respect to the equilibrium
value) depending on the lattice constant in KNbO$_3$
as calculated by LMTO-LDA (crosses),
LAPW-LDA (filled circles) and LAPW-GGA (open circles).
The parabolic fit is shown for each case.
Experimental lattice constant extrapolated to zero temperature
is indicated by a vertical line.}
\label{vol_fig}
\end{figure}

Fig.~\ref{vol_fig} shows the energy/volume curves as calculated
by LAPW in the LDA and in the GGA, and by LMTO in the LDA.
First two curves essentially reproduce previous results
by Singh\cite{sin95} aimed at the comparison of LDA and GGA.
The energy/volume curve generated now
with LMTO (crosses in Fig.~\ref{vol_fig}) practically coincides
with that obtained with the LAPW. Absolute energy values lie
by $\sim$0.9 Ry lower with LAPW, understandably due to more
complete basis.

\begin{figure}[t!] 
\centerline{\epsfig{file=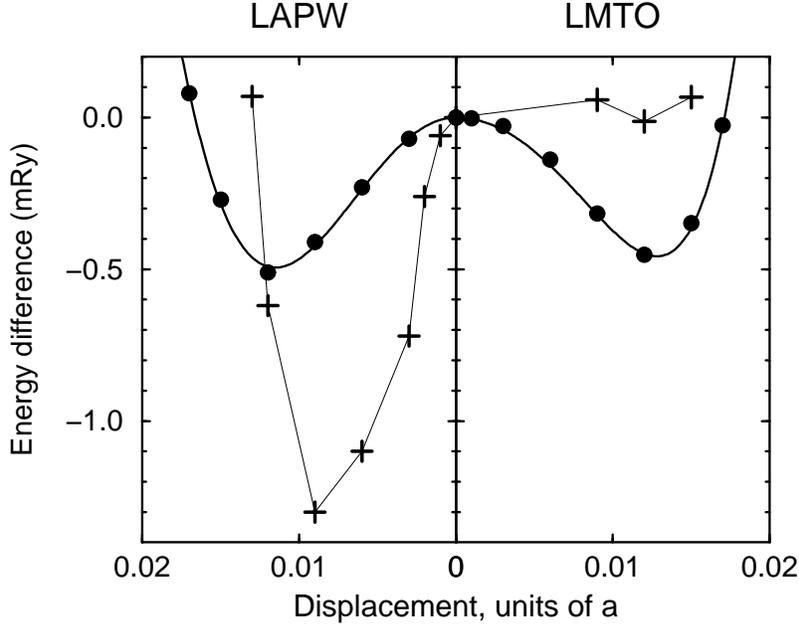,height=8.0cm}}
\vspace{10pt}
\caption{Total energy difference depending on atomic displacements
$\Delta_z(\mbox{K}) + 2\Delta_z(\mbox{Nb})$
as calculated by LAPW and LMTO for different values of $\Delta_z$
and different cutoffs in the charge density expansion.
Crosses: $G$=12.0 in LAPW and $N$=18 in LMTO;
dots: $G$=14.0 in LAPW and $N$=24 in LMTO
(essentially converged results). Polynomial fit is a guide to eye.}
\label{disp_fig}
\end{figure}

For the study of ferroelectric instability, we concentrated on
the displacement pattern compatible with the $t_{1u}$ TO phonon modes,
i.e. the $z$-displacements of K\mbox{(0 0 0)},
Nb\mbox{($\frac{1}{2} \frac{1}{2} \frac{1}{2}$)}
and O$_{\mbox{\tiny II}}$ \mbox{($\frac{1}{2} \frac{1}{2} 0$)} with respect to
two equivalent O$_{\mbox{\tiny I}}$ \mbox{($0 \frac{1}{2} \frac{1}{2}$)}
and \mbox{($0 \frac{1}{2} \frac{1}{2}$)} atoms.
In Fig.~\ref{disp_fig}, the energy differences are shown as function
of the displacement pattern which roughly corresponds to that
in the soft mode, ultimately resulting in the equilibrium
structure of tetragonal ferroelectric phase:
Nb is displaced twice farther as K relatively to the oxygen cage.
We found the calculated energy differences
to be extremely sensitive to the quality of the charge density
representation in the unit cell. In both methods we used, this expansion
is done by the fast Fourier transformation; in WIEN97, the magnitude
of the largest reciprocal-space vector $G$ is specified
whereas in the NFP-LMTO the number of divisions $N$ along
each unit cell edge for a real-space uniform grid has to be
explicitly provided.
For a perovskite, both cutoffs need to be relatively large
in order to achieve a convergency in this parameter.
As is seen in Fig.~\ref{disp_fig}, the value $G=10$ in LAPW
overestimates the ferroelectric instability
whereas the LMTO with $N=18$ finds yet no trace of the instability.
At $G=12$ and $N=24$, the trends are about comparable,
becoming even closer for $G=14$ and $N=32$.

\begin{table}[t!]
\caption{Calculated frequencies and eigenvectors of $\Gamma$-TO phonons
of the $t_{1u}$ symmetry from LAPW and LMTO}
\label{tab:phonon}
\begin{tabular}{cdddd@{\hspace*{0.8cm}}cdddd}
$\omega$ & \multicolumn{4}{c}{Eigenvectors} &
$\omega$ & \multicolumn{4}{c}{Eigenvectors} \\
(cm$^{-1}$) & K & Nb & O$_{\mbox{\tiny I}}$ & O$_{\mbox{\tiny II}}$ &
(cm$^{-1}$) & K & Nb & O$_{\mbox{\tiny I}}$ & O$_{\mbox{\tiny II}}$ \\
\tableline
\multicolumn{5}{c}{LAPW, Ref.~\protect\cite{sb92}} &
\multicolumn{5}{c}{LMTO, Ref.~\protect\cite{phonon}} \\
115$i$ &    0.04 & $-$0.60 & 0.40 &    0.56 &
203$i$ &    0.32 & $-$0.67 & 0.29 &    0.53 \\
 168   & $-$0.88 &    0.35 & 0.19 &    0.16 &
 193   & $-$0.81 &    0.12 & 0.36 &    0.27 \\
 483   &    0.03 & $-$0.09 & 0.46 & $-$0.75 &
 459   &    0.13 & $-$0.14 & 0.45 & $-$0.75
\rule[-2mm]{0mm}{0mm} \\
\multicolumn{5}{c}{LAPW, Ref.~\protect\cite{krak96}} &
\multicolumn{5}{c}{NFP-LMTO, present results} \\
197$i$ &    0.01 & $-$0.59 & 0.42 &    0.55 &
106$i$ &    0.07 & $-$0.61 & 0.45 &    0.48 \\
 170   & $-$0.88 &    0.37 & 0.18 &    0.15 &
 179   & $-$0.88 &    0.33 & 0.21 &    0.17 \\
 473   &    0.02 & $-$0.08 & 0.46 & $-$0.76 &
 518   &    0.03 & $-$0.03 & 0.41 & $-$0.81
\rule[-2mm]{0mm}{0mm} \\
\end{tabular}
\end{table}

As an additional test for the proper balance of the energetics of
different displacement patterns, we calculated the $\Gamma$-TO
phonons in the cubic phase of KNbO$_3$. The insufficient accuracy
of previous LMTO calculations \cite{phonon,ortho} manifested itself
as a noticeable deviation of the vibration eigenvectors from those
obtained by LAPW \cite{sb92,krak96}. Table \ref{tab:phonon}
shows the results of some preliminary calculations with the NFP code
(obtained with the 2d-order total energy fit over the results
for several combined displacements) in comparison with the LAPW
data. One can see that the correct displacement pattern within
the soft mode is now restored, and the overall agreement with
the LAPW eigenvectors is quite satisfactory.

\section*{Treatment of anharmonic vibrations}

With the total-energy fit generally available from first-principles
calculations, one may tend to extract some additional information
than is possible within the harmonic approximation. The treatment
of anharmonic effects in crystal is rather complicated (see, e.g.,
Ref.~\cite{anhar} for a review). In principle, the modes of different
symmetry and related to different {\bf q}-values couple beyond
the harmonic approximation. Nevertheless, in the study of ferroelectrics
there have been several attempts to single out any particular mode,
which is believed to be principally associated with anharmonic
effects, and to solve the vibrational Schr\"odinger equation
related to it. This has been done e.g. for LiTaO$_3$ and
LiNbO$_3$ by Inbar and Cohen \cite{inco_95} and by Bakker {\it et al.}
\cite{bakker} (for an empirical potential well, in the latter case)
as well as for a one-dimensional $A_2$ mode in orthorhombic
KNbO$_3$ by Postnikov and Borstel \cite{ortho}.
This approach was referred to as non-interacting anharmonic
oscillators \cite{inco_95}, meaning the oscillators related to
a particular $\Gamma$ TO-mode in crystal. It is assumed that the
displacement potential is separable into components with
different {\bf q}-dependence. Such separation is less valid
for several symmetry coordinates which mix already in the harmonic
approximation, therefore the solution of a multidimensional oscillator
problem is necessary in this case.
A straightforward treatment by, e.g., a finite-difference method
on a multidimensional grid rapidly becomes prohibitive with a
number of dimensions (see \cite{ortho}). Therefore, we propose
a scheme which uses the expansion in hyperspherical harmonics.
This approach is known in the calculation of vibration spectra
of three-atomic molecules \cite{3at_vib}, however, for an
{\it ad hoc} constructed system of variables. In the following,
we describe the formalism for an arbitrary number of symmetry
coordinates.

We start with an arbitrary convenient set of symmetry-adapted
displacement coordinates (see, e.g., Ref.\cite{ck90}):
$S_t=\sum_i B_{ti}x_i$
($x_i$ are conventional cartesian displacements),
which form a complete basis within a particular
irreducible representation, but do not need to be
orthonormal. The Schr\"odinger equation then acquires a form:
\begin{equation}
\label{Sch_eq}
 \left[- \frac{~\hbar^2}{2}\sum_{tt'}
     \frac{\partial}{\partial S_t}G_{tt'}
     \frac{\partial}{\partial S_{t'}} +
     V(\{S_t\})\right] \Psi = E \Psi,
\end{equation}
with the kinetic-energy matrix
$G_{tt'}=\sum_i B_{ti} m^{-1}_i B_{t'i}$.
Mixed derivatives can further be excluded
by the following orthogonalizing transformation:
$$
Q_t = \sum_{t'} \frac{X_{t't}}{\sqrt{\lambda^t}}\,S_{t'},
$$
where $X_{t't}$ is the $t$-th eigenvector, corresponding
to the eigenvalue $\lambda^t$, of the kinetic-energy
matrix. In $n$-dim. space of generalized coordinates $Q_t$,
we use spherical coordinates
$(r, \vartheta_1,\cdots,\vartheta_p, \varphi)$ for $p=n\!-\!2$,
$$
\begin{array}{r@{\,=\,}l}
Q_1     & r\cos\vartheta_1 \\
Q_2     & r\sin\vartheta_1\cos\vartheta_2 \\
\multicolumn{2}{c}{\cdots} \\
Q_{p+1} & r\sin\vartheta_1\sin\vartheta_2 \cdots
	  \sin\vartheta_{p-1}\sin\vartheta_p\cos\varphi \\
Q_{p+2} & r\sin\vartheta_1\sin\vartheta_2 \cdots
	  \sin\vartheta_{p-1}\sin\vartheta_p\sin\varphi \\
\end{array}
$$
$$
0\le\vartheta_j\le\pi\quad(j=1,2,\cdots,p),\quad 0\le\varphi\le 2\pi \,.
$$
There are $(2N\!+\!p)(N\!+\!p\!-\!1)!/(p!\,N!)$ harmonic polynomials
of degree $N$ numbered by $n\!-\!1$ integers $m_0,\cdots,m_p$ such that
$N = m_0 \ge m_1\ge\cdots\ge |m_p|\ge 0$, $m_p\!=\!\pm |m_p|$.
The explicit form of the polynomials of is the following:
$$
H(N,m_1,..,m_p;\,Q_1,..,Q_{p+2}) =
r^N Y^{m_p}_{N,m_1,..,m_{p-1}}
(\vartheta_1,..,\vartheta_p,\varphi)\,.
$$
The hyperspherical harmonics are chosen either as complex functions
$$
Y^{m_p}_{m_0,.,m_{p-1}}\!
(\!\vartheta_1,\!.,\!\vartheta_p,\!\varphi\!)\!=\!
e^{im_p\varphi}\!\prod_{k=0}^{p-1}\!
(\sin\!\vartheta_{k+1}\!)^{m_{k+1}}
C^{m_{k+1}\!+\!\frac{p-k}{2}}_{m_k-m_{k+1}}
\!(\!\cos\!\vartheta_{k+1}\!)\;,
$$
or in the real form, $\sim\!\cos{m_p}$ for $m_p\ge 0$
and $\sim\!\sin{m_p}$ for $m_p<0$.
$C^p_n(z)$ are Gegenbauer polynomials \cite{gegen,takeu},
generated by the following recursion:
$$
C^p_0(z) = 1; \quad\quad
C^p_1(z)=2pz; \quad\quad
(n+1)\,C^p_{n+1}(z)=2(n+p)\,z\,C^p_n(z)-(n+2p-1)\,C^p_{n-1}(z)\;.
$$
$Y^{m_p}_{m_0,.,m_{p-1}}(\vartheta_1,\!.,\!\vartheta_p,\varphi)$,
orthogonal on a unit sphere,
are eigenfunctions of the multidimensional Laplace operator:
$$
\Delta Y = -\,\frac{m_0(m_0+p)}{r^2}\,Y\,.
$$
With the potential and the wave function expanded in hyperspherical
harmonics
\begin{eqnarray*}
V(\vec{Q})&=&\sum_{m_0,..m_p} \!V_{m_0,..,m_p}(r)
Y_{m_0,..,m_p}(\vartheta_1,\!..,\vartheta_p,\varphi)\,, \\
\Psi(\vec{Q})&=&\sum_{m'_0,..m'_p} \!R_{m'_0,..,m'_p}(r)
Y_{m'_0,..,m'_p}(\vartheta_1,\!..,\vartheta_p,\varphi)\,,
\end{eqnarray*}
the multidimensional Schr\"odinger equation (\ref{Sch_eq}) transforms
into a system of
$\sum_{N=0}^{N_{\mbox{\tiny max.}}}\!(2N+p)(N\!+\!p\!-\!1)!/(p!\,N!)$
coupled 1-dimensional equations:
\begin{eqnarray}
\nonumber
-\frac{\hbar^2}{2}\,\frac{1}{r^{p+1}}\,\frac{d}{dr}\!\left[
r^{p+1}\frac{dR_{[m]}(r)}{dr}\right] +
\frac{m_0(m_0+p)}{r^2}\,R_{[m]}(r) \;+ \\
+ \!\sum_{[m''][m']}
\!\!V_{[m'']}(r)R_{[m']}(r)\!
\int\!\!Y_{[m'']}Y_{[m']}Y_{[m]}d\,\Omega = E\,R_{[m]}(r)\,.
\label{R_equa}
\end{eqnarray}
The expansion of the potential (provided in a polynomial form
by a fit to total-energy values) is finite whereas for
the wave function a cutoff value $N_{\mbox{\tiny max.}}$ has to
be introduced.

As a practical example of this approach, we considered the solution
of a 3-dimensional oscillator problem corresponding to the vibration pattern
within the $t_{1u}$ mode in cubic KNbO$_3$. For the symmetry coordinates
as discussed above, we included the 4th power of the Nb displacement
into the total energy fit. (For real applications, one should of course
consider some other degrees of freedom beyond the harmonic approximation).
The system of coupled equations (\ref{R_equa})
was solved by a finite difference method, with 50 points
in the equidistant radial mesh from up to $r=5.0$ where a boundary
condition $R(r)=0$ was imposed on radial wavefunctions (this scheme
may be somehow refined in more precise calculations, incorporating
a nonuniform mesh).
For the maximal degree of polynome $N=8$ in the wave function expansion,
the energy difference between two lowest oscillator levels
practically converged to 70 cm$^{-1}$. The convergence is more slow
for higher levels.

\section*{Conclusions}

We compare in the present paper the results obtained
for the ferroelectric instability in KNbO$_3$ with two methods,
LAPW and LMTO, both of which have been applied to this system before
but apparently never underwent a thorough comparison with
the, as far as possible, identical calculation setup. The result
of this comparison is that not only the energy/volume curves are
identical in the LDA, but the description of the ferroelectric
instability, involving equilibrium displacements of $\sim$0.1 a.u.
and energy differences of $\sim$0.5 mRy, is practically identical
by both schemes, provided the sufficient accuracy in the description
of the charge density variations over the unit cell is guaranteed.
The LAPW method provides understandably lower absolute values
of the total energy, but the new formulation of LMTO has the advantage
of much more compact basis set (about 10 times smaller than that of LAPW)
and some resources to expand the basis somehow for even better
controllable accuracy without running into numerical problems of
overcompleteness. As a useful tool for the analysis of total
energy data obtained from any first-principle calculation, we describe
the scheme to solve the multi-dimensional vibrational Schr\"odinger equation
in the approximation of non-interacting anharmonic oscillators.
The preliminary results for the lowest energy difference are presented
for KNbO$_3$.\\*[0.8cm]
\centerline{\bf Acknowledgments}\\*[0.3cm]
We thank M.~Methfessel for providing us with his new version
of the LMTO code and for numerous helpful discussions.
We are grateful to K.~Schwarz and P.~Blaha for making
the WIEN97 code available to us.
C.~O.~Rodriguez valuably contributed in working
discussions, e.g., on applying the LAPW method to KNbO$_3$.
A.P. is grateful to D.~Vanderbilt for stimulating discussions
on lattice dynamics.
Financial support of the Deutsche Forschungsgemeinschaft (SFB 225)
is gratefully acknowledged.

\end{document}